\documentclass[two column,prb,showpacs,multicol,amsmath,amssymb]{revtex4}
\usepackage[dvips]{graphicx}
\usepackage{graphicx}
\usepackage{dcolumn}
\usepackage{bm}
\usepackage{graphics}
\usepackage{epsfig,color}

\begin{document}
\title[A. Ahmadi Fouladi ]{Effects of impurity on tunnel magnetoresistance in a ferromagnetic electrode/carbon nanotube/ferromagnetic electrode junction}
\author{A. Ahmadi Fouladi, J. Vahedi and M. Soleymani}
\address{ Department of Physics, Sari Branch, Islamic Azad University, Sari, Iran.}
\date{\today}

\begin{abstract}
Effects of impurity on the spin-dependent transport in a single wall carbon nanotube spin-valve, as ferromagnetic electrode/carbon nanotube/ferromagnetic electrode model junction is numerically investigated. Using a generalized Green's function method and the Landauer-B\"uttiker formalism, the impurity conditions are determined by randomly substitution of carbon atoms in the honeycomb carbon nanotube lattice by nitrogen and boron atoms. We have found that transport characteristics, including the spin-dependent current and tunnel magnetoresistance are strongly influenced by the impurity effects. We think that the results of the present report could be useful for designing the future spintronic devices.
\end{abstract}

\maketitle
\section{Introduction}
In recent years nanotechnology has attracted attention of numerous researchers due to its vast potential applications \cite{1,2,3,4,5,6,7,8,9,10,11,12,13,14,15,16,17,18,19,20,21,22}. Spintronics is one of the most attractive research areas of nanotechnology. The development of spintronic devices based on the tunnel magnetoresistance (TMR), giant magnetoresistance (GMR) and spin-valve effects has changed magnetic memory applications.
In the conventional spin-valve geometry, a non-magnetic spacer, which controls the total resistance of the device, the spin polarization and transport, is sandwiched between two magnetic contacts. Among many suggestions for possible non-magnetic spacers, organic materials provide desirable properties like flexibility, inexpensive material and production methods have attracted investigations in the organic electronics \cite{1,2,3,4,5,6,7,8,9,10,11,12}. Besides, the spin-orbit coupling and hyperfine interactions in the organic materials are very weak so that the spin memory can be as long as a few seconds. Therefore the spin-flip process during transport can be neglected \cite{3,9}. Such properties make them ideal for spin-polarized transport applications. Among many types of organic materials, the carbon nanotube (CNT) is a promising candidate as a component in spintronic devices. Spin-polarized transport through the CNT contacted by ferromagnetic leads has also been recently investigated \cite{1,12,13,14,15,16,17}.
\par
\begin{figure}
\includegraphics[width=0.9\columnwidth]{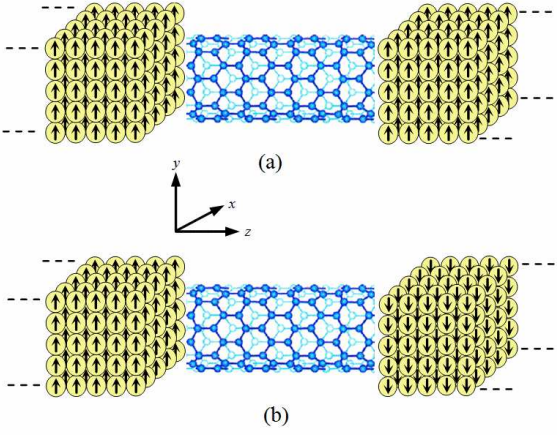}
\caption {A schematic FM/CNT/FM junction model for (a) parallel and (b) anti-parallel spin alignments.}
\label{p1}
\end{figure}

In the other hand, understanding of the consequence of doping on the electronic efficiency of devices based on low-dimensional carbon-structures is a key-point for the future development of carbon based nanoelectronics and to the finding of novel functionalities. Namely, since the electronic properties of CNTs are strongly related to the delocalised electron system, clearly any modification of the CNTs will influence these properties. Therefore, by a suitable choice of the type of modification the electronic properties of CNTs can be intentionally tuned. The tuning of electronic properties is also referred to as impurity doping the CNTs. Also the substitution of carbon atoms in the honeycomb lattice by atoms with a different number of valence electrons in general introduces additional states in the density of states. Nitrogen (N) and boron (B) are the natural choice for doping since they differ only by one in its number of valence electrons compared to carbon atoms.
There have been several theoretical \cite{23,24,25} studies on the doping of CNTs by B and N atoms. Also several authors \cite{26,27,28,29,30,31,32,33,34,35} have used different experimental methods to prepare doped CNT with nitrogen and boron. In this paper, we have numerically investigated the coherent spin-dependent transport through (5,0) zigzag CNT with 1.56 nm length attached to ferromagnetic 3-dimensional leads (FM/CNT/FM junction), as depicted in Figs. \ref{p1}(a) and \ref{p1}(b) for parallel and anti-parallel spin alignments, respectively. Using the Green's function method and relying on the Landauer-B\"uttiker formalism, we have studied the influence of impurities on spin-dependent transport properties and TMR in a FM/CNT/FM model junction. In this work the impurity conditions are determined by randomly replacing carbon atoms in the honeycomb CNT lattice with nitrogen and boron atoms.

\section{Computational scheme} \label {sec:1}
A generalized Hamiltonian for the FM/CNT/FM junction can be expressed as;
\begin{equation}
\emph{H}=H_{CNT}+H_{L}+H_{R}+H_{C}\,.
\label{e1}
\end{equation}
The first term is the Hamiltonian for the CNT within the non-interacting picture given by
\begin{eqnarray}
\emph{H}_{CNT}=\sum_{{n},{\sigma}}\varepsilon_{n}d_{{n},{\sigma}}^{\dag}d_{{n},{\sigma}}-\sum_{{<n,m>},{\sigma}}t_{C-C}(d_{{n},{\sigma}}^{\dag}d_{{m},{\sigma}}+H.c.).\
\label{e2}
\end{eqnarray}
where $n$ runs over the $\pi$-orbitals of a carbon atom on the CNT. The operator $d_{{n},{\sigma}}^{\dag}(d_{{n},{\sigma}})$ creates (annihilates) an electron with spin $\sigma\equiv\uparrow or\downarrow$ at site $n$. $\varepsilon_{n}$ is the on-site energy of the carbon atom and $t_{C-C}$ is the nearest neighbour electron hopping integral. Within the non-interacting picture, the single-band tight binding Hamiltonian of the left (right) ferromagnetic electrode can be written as;
\begin{eqnarray}
\emph{H}_{\beta}=\sum_{i_{\beta},{\sigma}}\{(\varepsilon_{0}&-&{\mathbf\sigma}.{\bf J_{\beta}})c_{i_{\beta},{\sigma}}^{\dag}c_{i_{\beta},{\sigma}}\nonumber\\
      &-&t(c_{i_{\beta},{\sigma}}^{\dag}c_{i_{\beta}+1,{\sigma}}
             +c_{i_{\beta}+1,{\sigma}}^{\dag}c_{i_{\beta},{\sigma}})\}.
\label{e3}
\end{eqnarray}
where $c_{i_{\beta},{\sigma}}^{\dag}(c_{i_{\beta},{\sigma}})$ denotes the creation (annihilation) operator of an electron with spin $\sigma$ at site $i$ in the electrode $\beta{(=L or R)}$. $t$ and $\varepsilon_{0}$ are the nearest neighbour hopping integral and the on-site energy, respectively. Here $-{\bf\sigma}.\bf J_{\beta}$ is the internal exchange energy and $\bf J_{\beta}$ denotes the molecular field at site $i_{\beta}$.
The last term of Eq. \ref{e1}, $H_{C}$  denotes the coupling between CNT and FM electrodes
\begin{eqnarray}
\emph{H}_{C}=\sum_{{n},{\sigma}}\sum_{{i},{\sigma}}t_{c(n,{\sigma},i)}(c_{{i},{\sigma}}^{\dag}d_{{i},{\sigma}}+H.c),\
\label{4}
\end{eqnarray}
where the matrix elements $t_{c(n,\sigma,i)}$ represent the coupling strength between CNT, and FM electrodes are taken to be $t_{c}$. For a complete system, i.e., CNT with two FM electrodes, the spin-dependent Green's function can be written as;
\begin{equation}
G_{\sigma}(E)=\big[E1-H_{CNT}-\Sigma_{L,\sigma}-\Sigma_{R,\sigma}]^{-1},
\label {e5}
\end {equation}
$\Sigma_{L,\sigma}$ and $\Sigma_{R,\sigma}$ are the self-energy matrix due to coupling of CNT to the left and right electrodes respectively and given by
\begin {equation}
\Sigma_{\beta,\sigma}(E)=\tau_{CNT,\beta}\,\, g_{\beta,\sigma}\,\,\tau_{\beta,CNT},
\label{e6}
\end{equation}
where $\tau$ is the hopping matrix that couples CNT to the FM electrodes, and $g_{\beta,\sigma}$ is the surface Green's function of FM electrodes which is given by \cite{36};
\begin{equation}
g_{\beta,\sigma}(n,m;z)=\sum_k \frac{{\psi}_k({\bf r}_n){\psi}_k^*({\bf r}_m)}{z-\varepsilon_{0}+{\bf\sigma}.{\bf J}_{\beta}-E(\bf k)},
\label{e7}
\end{equation}
where ${\bf r}_n \equiv (x_n,y_n,z_n)$, ${\bf k} \equiv (l_x,l_y,k_z)$ ,  $z=E+i\eta$ ,
\begin{eqnarray}
{\psi}_k({\bf r}_n)&=&\frac {2\sqrt{2}}{\sqrt{(N_x+1)(N_y+1)N_z}} \sin(\frac{l_x x_n \pi}{N_x+1})\nonumber\\
&&\times\sin(\frac{l_y y_n \pi}{N_y+1})\sin({k_z z_n}),
\label{e8}
\end{eqnarray}
and
\begin{eqnarray}
E({\bf k})=2t\big[\cos(\frac{l_x \pi}{N_x+1})+\cos(\frac{l_y \pi}{N_y+1})+\cos({k_z a})\big].
\label{e9}
\end{eqnarray}
Here, $l_{x,y}(=1,...,N_{x,y})$ are integers, $N_\xi$ with $\xi=x,y,z$ is the number of lattice sites in the $\xi$ direction and $k \in [-\pi/a ,\pi/a]$.
The number of atoms at the cross-section of FM electrodes are taken to be $N_x=N_y =5$. We assume that only the central atom at the electrode cross-section connected to the one carbon atom at the end of CNT lattice. The broadening matrix $\Gamma_{L,\sigma}(\Gamma_{R,\sigma})$ can be calculated through the expression,
\begin{equation}
\Gamma_{\beta,\sigma}=-2 Im(\Sigma_{\beta,\sigma}).\
\label{e10}
\end{equation}
Neglecting the spin-flip scattering effects, one may consider transport of spin-up and spin-down electrons separately. The spin-dependent transmission probability of an electron in the CNT can be calculated using the following relation \cite{37}:
\begin{equation}
T_{\sigma}(E)=Tr(\Gamma_{L,\sigma}G^r_\sigma \Gamma_{R,\sigma}G^a_\sigma).
\label{e11}
\end{equation}
Where $G^r_\sigma$ and $G^a_\sigma$ are the retarded and advanced Green's function. Based on non-equilibrium Green's function method, the spin-dependent current as a function of the applied bias voltage in the low-bias limit can be computed in the framework of the Landauer-B\"uttiker formula \cite{37}:
\begin{equation}
I_{\sigma}(V_{a})=\frac{e}{h}\int_{-\infty}^{+\infty}T_{\sigma}(E)\big[f_{L}-f_{R}\big] dE ,
\label{e12}
\end{equation}
where $f_{L(R)}=f(E-\mu_{L(R)})$ is the Fermi distribution function at the left (right) electrode with chemical potential $\mu_{L(R)}=E_{F}\pm\frac{eV_{a}}{2}$ and Fermi energy $E_{F}$. For the sake of simplicity, here we have assumed that the total voltage is dropped across CNT/electrode interface, and this assumption does not highly affect the qualitative aspects of current-voltage characteristics. In fact, the electric field inside the CNT, especially for short length CNT, seems to have an insignificant effect on the $I-V$ characteristics. On the contrary, for longer CNT and higher bias voltages, the electric field inside the CNT may play a more considerable role depending on the structure of the CNT, but yet the effect is very small \cite{38}. The tunnel magnetoresistance (TMR) can be defined as a relative change in the current of the system when magnetization of two FM electrodes switch between parallel $(\mathit{P})$ and anti-parallel $(\mathit{AP})$ configurations, hence: $TMR \equiv \frac {I_P-I_{AP}}{I_P}$. In fact the TMR is associated with asymmetry of the density of states for two spin channels in the FM electrodes. 
\section{Results and discussion}
Based on the formalism described in section 2, we represent the results of numerical calculations. Tight-binding parameters for FM electrodes chosen to be $\varepsilon_{0}=3\,eV$, $t=1\,eV$ and ${|\bf J_{\beta}|}=1.25 \,eV$. In our calculation, the on-site energy of C, N and B atoms are assumed to be 0, -2.50 and +2.33 eV, respectively \cite{39,40}. The nearest neighbour electron hopping integrals are $t_{C-C}=-3\,eV$, $t_{C-B}=-2.7\,eV$ and $t_{C-N}=-3.14\,eV$. As a reference energy, the Fermi energy of FM electrodes is set $E_F=0$ and the coupling between CNT and two FM electrodes set as $t_c=0.3\,eV$. Magnetization direction in the left ferromagnetic electrode is fixed in the $+y$ direction, while the right electrode is free to be flipped into either the $+y$ or $-y$ direction by an external magnetic field. For parallel $(\mathit{P})$ alignment of magnetization in the FM electrodes, spin-up and spin-down electrons encounter a symmetric structure, while for anti-parallel $(\mathit{AP})$ alignment these electrons encounter an asymmetric structure. A low temperature of $T=11\,K$ is taken to avoid spin flip in the electron transport progress \cite{2}.

\begin{figure*}
\includegraphics[width=1.5\columnwidth]{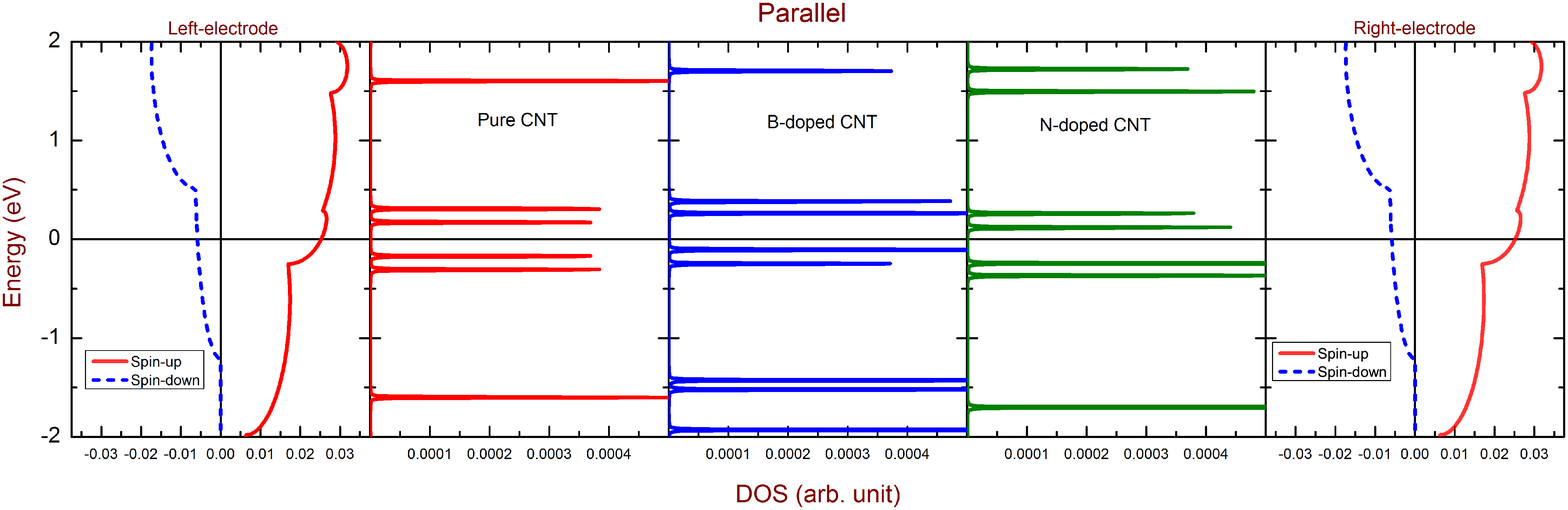}\\
\includegraphics[width=1.5\columnwidth]{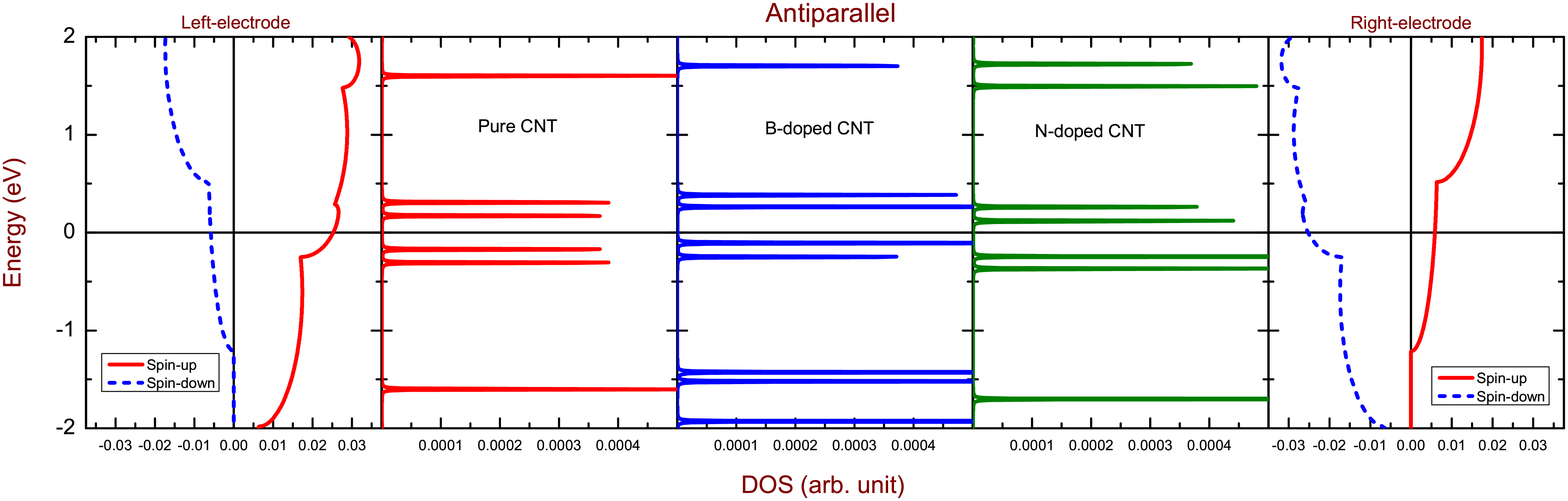}
\caption {Surface density of states of the isolated FM electrodes and density of states of the isolated CNT for the pure case and randomly doped B and N atoms with $8\%$ concentrations for parallel (top panel) and anti-parallel (bottom panel) spin alignments.}
\label{p2}
\end{figure*}

In Fig.\ref{p2}, we have plotted surface density of states (SDOS) of the isolated FM electrodes and density of states (DOS) of the isolated CNT for pure and randomly doped B and N atoms with $8\%$ concentrations in both parallel and anti-parallel configurations. The discreteness of molecular levels is observed from DOS graphs. Furthermore, Fig.\ref{p2} shows that dopant shifts HOMO and LUMO levels away from the Fermi energy. Therefore, in the presence of dopants, symmetry of DOS about the Fermi energy (for pure case) is broken. The shift of HOMO and LUMO levels depends on concentration of B and N atom doping. B doping (N doping) of CNT shifts HOMO and LUMO levels towards the higher (lower) energies \cite{41}. Consequently, HOMO-LUMO energy gap will change.

\begin{figure*}
\includegraphics[width=1.1\columnwidth]{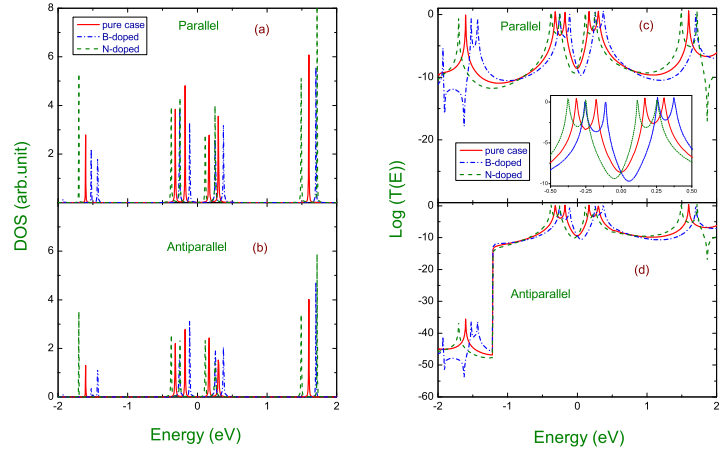}
\caption{Density of states ((a) and (b)) and logarithmic scale of transmission function ((c) and (d)) versus energy for the FM/CNT/FM junction in $\mathit{P}$ and $\mathit{AP}$ configurations for the pure case and randomly doped B and N atoms with $8\%$ concentrations.}
\label{p3}
\end{figure*}

We then calculated DOS and logarithmic scale of transmission function for FM/CNT/FM junction for $\mathit{P}$ and $\mathit{AP}$ configurations, which are presented in Fig. \ref{p3}. As it can be seen, when isolated CNT contacted to FM electrodes, the contribution of the electronic levels of the electrode surface increases DOS. However, we can see a remarkable difference between DOS of the $\mathit{P}$ and $\mathit{AP}$ configurations, which can effect on transmission function as shown in Figs. \ref{p3}(a) and \ref{p3}(b).

The resonance peaks in transmission probability are associated with eigenenergies of CNT. When electrons go from the left to right FM electrode one through   CNT, electron waves propagating along different branches of CNT honeycomb lattice may suffer a relative phase shift between themselves. Consequently, there might be constructive or destructive interference due to the superposition of the electronic wave functions along various pathways. Therefore, transmission probability changes. Also, we observe some anti-resonant states appear in transmission probability in Figs.\ref{p3}(c) and \ref{p3}(d). These anti-resonant states are related to the quantum interference effect. In general, transmission probabilities are higher for  $\mathit{P}$ alignment than  $\mathit{AP}$ alignment. This difference is associated with the asymmetry of  SDOS of the FM electrodes for spin-up and spin-down electrons as shown in Fig.\ref{p2} and quantum tunnelling phenomenon through a CNT.In the presence of N and B atoms, with shifting HOMO and LUMO levels, the position of resonance peaks changes (inset of Fig.\ref{p3} (c)).

\begin{figure*}
\includegraphics[width=0.8\columnwidth]{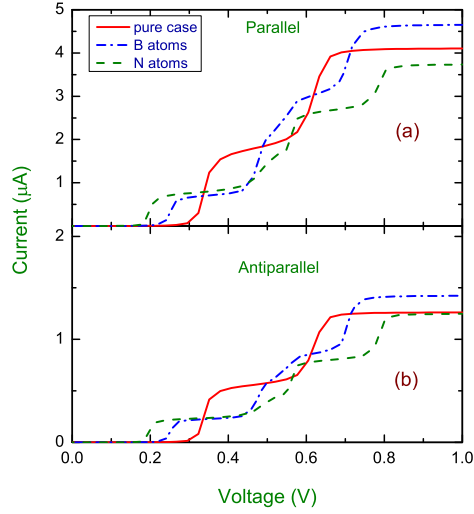}
\caption{Current-voltage ($I-V$) characteristics for (a) $\mathit{P}$ and (b) $\mathit{AP}$ configurations for the pure case and randomly doped B and N atoms with $8\%$ concentrations}
\label{p4}
\end{figure*}

In order to provide a deep understanding of spin-dependent transport, we have plotted current-voltage characteristics of FM/CNT/FM system for (a) $\mathit{P}$ and (b) $\mathit{AP}$ configurations in Fig.\ref{p4}. It is observed that the presence of impurities have a profound effect on the current amplitude. Threshold voltage changes in the presence of B and N atoms, owing to altering the energy gap. By decreasing energy gap, the nearest molecular levels to the gap is less separated from the Fermi level, and smaller voltage is needed for turning current on. In $\mathit{P}$ configuration the majority (minority) electrons with spin-up (spin-down) in the left FM electrode move into the majority (minority) states in the right FM electrode by tunnelling through CNT. For $\mathit{AP}$ configuration the majority (minority) electrons with spin-up (spin-down) in the left electrode move into the minority (majority) states in the right FM electrode. Therefore, total current amplitude in $\mathit{P}$ alignment is higher than $\mathit{AP}$ alignment. 

\begin{figure*}
\includegraphics[width=0.8\columnwidth]{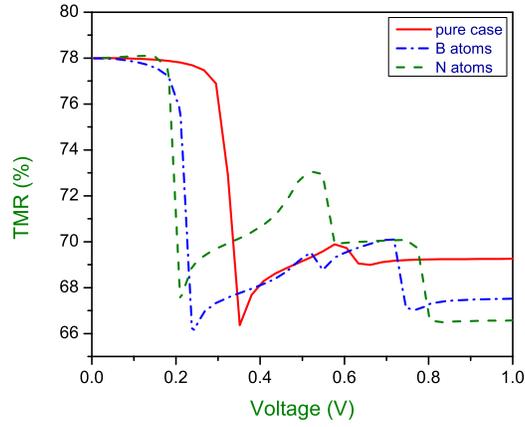}
\caption{TMR ratio as a function of an applied bias for the pure case and randomly doped B and N atoms with $8\%$ concentrations}
\label{p5}
\end{figure*}

In Fig.\ref{p5} we have depicted TMR ratio as a function of applied bias voltage in the absence and presence of impurities. As illustrated, TMR ratio shows maximum value ($78\%$) for lower bias voltages. By increasing applied voltage, TMR shows a sharp drop and after that it varies slowly. In the low bias voltages for  maximum value of TMR, there is no molecular level available in the gap region of CNT so that electrons can tunnel between the chemical potential of the left and right FM electrodes. Then, in the low-voltage regime current is small and TMR ratio is large. When bias voltage increases, the electrochemical potentials in electrodes are shifted gradually, and some of the energy states are pushed up between the chemical potential of the right and left electrodes and  current of tunnelling through junction increases remarkably. Our results are qualitatively in agreement with the experimental measurements \cite{41}. Moreover, because of the different energy gaps in the absence and presence of impurity, we can see that the first sharp decrease of TMR happens for different values of bias voltage. As a result one may tune TMR ratio with applied voltage and also impurities. 

\begin{figure*}
\includegraphics[width=0.8\columnwidth]{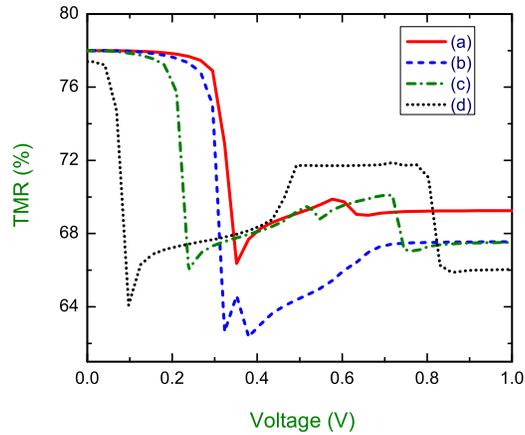}
\caption{TMR ratio as a function of an applied bias for (a) pure case and randomly doped B atom with (b) $4\%$, (c) $8\%$ and (d) $12\%$ concentrations.}
\label{p6}
\end{figure*}

In Fig.\ref{p6}, we have plotted TMR as a function of applied bias voltage for  different amount of B atoms as an impurity concentration. Because of the different HOMO-LUMO gap of CNT in the presence of different amounts of impurities, we can see that the decrease of TMR occurs for various values of bias voltage. Also for higher amount of  impurity ($12\%$), maximum value of TMR ratio decreases.
\section{Conclusions}
To summarize, based on Landauer-B\"uttiker formalism, we have studied the effects of impurity on spin-dependent transport through a single wall carbon nanotube (CNT) attached to ferromagnetic (FM) 3-dimensional leads as FM/CNT/FM model junction. We choose (5,0) zigzag CNT and study effects of randomly replaced carbon atoms in the honeycomb CNT lattice with nitrogen and boron atoms. Our results indicate that spin transport characteristics, including the spin-dependent current-voltage characteristics, the transmission probability and tunnel magnetoresistance (TMR) are strongly influenced by type of impurity as well as its concentration. Results show that the existence of impurities can change the HOMO-LUMO energy gap. Consequently, current amplitude and threshold voltage undergo changes. Finally, introduction of impurity can shift the first sharp decreasing point of TMR with decreasing the maximum value of TMR ratio.

\newpage

\end{document}